# Quantum Codes for Controlling Coherent Evolution

Yehuda Sharf[1], Timothy F. Havel[2], and David G. Cory[1,*]

[1]*Dept. of Nuclear Engineering, Massachusetts Institute of Technology, Cambridge, MA 02139, USA*
[2]*BCMP, Harvard Medical School, 240 Longwood Ave., Boston, MA 02115, USA*

MARCH 27, 2000

## ABSTRACT

Control over spin dynamics has been obtained in NMR via coherent averaging, which is implemented through a sequence of RF pulses, and via quantum codes, which can protect against incoherent evolution. Here, we discuss the design and implementation of quantum codes to protect against coherent evolution. A detailed example is given of a quantum code for protecting two data qubits from evolution under a weak coupling (Ising) term in the Hamiltonian, using an ``isolated'' ancilla, which does not evolve on the experimental time scale. The code is realized in a three-spin system by liquid-state NMR spectroscopy on 13C-labelled alanine, and tested for two initial states. It is also shown that for coherent evolution and isolated ancillae, codes exist that do not require the ancillae to initially be in a (pseudo-) pure state. Finally, it is shown that even with non-isolated ancillae quantum codes exist which can protect against evolution under weak coupling. An example is presented for a six qubit code that protects two data spins to first order.

* To whom correspondence should be addressed.
NW14-2217, Massachusetts Institute of Technology, Cambridge MA 02319
Phone: (617) – 253 3806
Fax: (617) 253-5405
E-mail: dcory@mit.edu

## INTRODUCTION

Quantum error correction (QEC) was developed to protect a quantum state from decoherence (1, 2, 3, 4, 5). To do this, the arbitrary state of the quantum bits, or qubits, of interest is encoded in a larger Hilbert space composed from additional qubits, called ancilla, initially prepared in a pure state. The encoded system is then allowed to evolve and interact with the environment. QEC codes are designed so that, after applying the inverse encoding, the state of the system is mapped into orthogonal subspaces according to a pre-selected key of errors such as bit flips, phase flips, or in general a complete set of operators for decoherence. A further correction step recovers the original state of the qubits of interest. The correction relies on the final state of the ancilla, which provides information about the error syndrome and thus the required correction steps. The correction step places the entropy in the ancilla, and this is then removed from the system of interest by tracing over (i.e. "discarding") the ancilla.

Experimental demonstrations of quantum error correction have been reported using liquid-state NMR (6, 7, 8). These studies exploited the high degree of coherent control provided by NMR to implement QEC codes and to test their performance under different models of induced decoherence. Moreover, these studies provide strong evidence that pseudo-pure states reproduce the dynamics of pure states, and that quantum coding may be used to validate theoretical decoherence mechanisms as well as to provide detailed information on correlations in the underlying relaxation dynamics (8).

Control over coherent spin dynamics is generally obtained in NMR by coherent averaging using pulse symmetries to select specific interaction frames. Here we generalize quantum error correction to the case of coherent evolution. As was recently argued, physically realistic models of quantum computers require the ability to correct correlated two qubit errors in addition to uncorrelated one qubit errors. This is achievable by error correction (9) and/or by protecting the data in decoherence free subspaces (10,11). A potential advantage of quantum coding over average Hamiltonian theory is that coherent control is attainable even though the individual spins in the system can not be addressed. This feature is particularly relevant to quantum information processing in quantum systems subjected to incomplete coherent control that leaves parts of the internal Hamiltonian unwillingly active.

To devise a quantum code it is necessary to know the symmetry of the coherent or decoherent evolution so that a class of possible errors can be defined. Spin dynamics are conveniently described in Liouville space by the master equation in the rotating frame (12)

$$\frac{d\rho}{dt} = -i\hat{\hat{H}}\rho + \hat{\hat{\Gamma}}(\rho - \rho_0) \qquad [1]$$

where $\hat{\hat{L}}$ is the coherent evolution superoperator, and $\hat{\hat{R}}$ is the relaxation superoperator. The solution of the set of differential equations [1] can be written as a time evolution superpropagator $\hat{\hat{T}}$, so that if the spin system is initially in a state $\rho(0)$, then

$$\rho(t) = \hat{\hat{T}}\rho(0).\qquad[2]$$

According to quantum information theory complete coherent control can be obtained using a small set of logic gates, called a universal set (13). In NMR, this degree of coherent control is attainable, as implied by average Hamiltonian theory, via selective RF excitations and free evolution delays under spin coupling (14). An example that can be considered as a benchmark for coherent control is the ability to turn off the internal Hamiltonian, so that $\hat{\hat{L}}$ becomes zero and the free evolution superpropagator is close to the identity (8, 15). However, if control over the quantum system is incomplete coherent evolution under the residual internal Hamiltonian will alter the state of the system. Under these conditions, coherent evolution due to the residual free evolution superoperator $\hat{\hat{L}}$ is regarded as a potential source of errors just as incoherent effects due to the relaxation superpropagator $\hat{\hat{R}}$. Nevertheless, quantum error correction is applicable regardless of the source of the error and therefore can protect against coherent evolution.

## PROTECTING TWO SPINS AGAINST EVOLUTION UNDER A $\sigma_z\sigma_z$ TERM USING AN ISOLATED ANCILLA

In the following we consider a simple example of a quantum code for suppressing coherent evolution. Here, the lack of complete control is expressed by the presence of a bilinear term $\sigma_z\sigma_z$ in the Hamiltonian. The state of two data spins is encoded in a Hilbert space including an additional ancilla spin, so that the state of the data spins can be restored. This example demonstrates the theory and practice of the elementary features of quantum coding such as preparation of pseudo-pure states, encoding, decoding, correction and tracing over the ancilla. It is simple enough to be analyzed both in terms of the standard computational basis as well as in the product operator formalism.

Consider a weakly coupled three spin system, the internal Hamiltonian of which is

$$H_{int} = \tfrac{1}{2}\omega_1\sigma_z^1 + \tfrac{1}{2}\omega_2\sigma_z^2 + \tfrac{1}{2}\omega_3\sigma_z^3 + \tfrac{\pi}{2}J_{12}\sigma_z^1\sigma_z^2 + \tfrac{\pi}{2}J_{13}\sigma_z^1\sigma_z^3 + \tfrac{\pi}{2}J_{23}\sigma_z^2\sigma_z^3 \qquad[3]$$

where $\omega_i$ are frequency shifts and $J_{kl}$ are coupling constants. The condition for weak coupling is $|J_{kl}| << \omega_{kl} \quad |\omega_k - \omega_l|$ for all $k$ and $l$.

The code is designed to protect the $\tfrac{\pi}{2}J_{12}\sigma_z^1\sigma_z^2$ term of the Hamiltonian (Eq. [3]). Therefore during the "error period" the unitary evolution propagator, assuming for simplicity that no relaxation takes place, is given by

$$U(t) = e^{-i\frac{\pi}{2} J_{12}\sigma_z^1\sigma_z^2 t} = \cos(\pi J_{12}t/2) - i\sigma_z^1\sigma_z^2 \sin(\pi J_{12}t/2). \quad [4]$$

It is assumed that the third spin is not coupled, or very weakly coupled to the other spins so that in the time scale of the experiment $\pi J_{ai}t << 1$ for all coupling constants. The third spin is therefore unaffected by the error, and therefore deserves the term "isolated" ancilla.

Ancilla qubits are necessary to provide a Hilbert space large enough to map all the errors into orthogonal subspaces. Since there is only a single error in our model, it is easy to see that a single ancilla provides sufficient encoding capacity. At the end of the decoding the ancilla will be either in the state $|0\rangle$ if no error has occurred, or in the state $|1\rangle$ in the case of a $\sigma_z^1\sigma_z^2$ error.

The code

A simple code that protects the data qubits against $\sigma_z^1\sigma_z^2$ errors is given in Fig. 1. It consists of controlled-NOT (c-NOT) gates as well as Hadamard gates. A c-NOT gate is a two-qubit operation that flips the state of one qubit conditional on the other being in the state $|1\rangle$. Conditional rotations are conveniently expressed in terms of geometric algebra (16). For example, the c-NOT$_{AB}$ gate, which flips spin B when spin A is in the state $|1\rangle$, has the form

$$\text{c-NOT}_{AB} = \sigma_x^B E_-^A + E_+^A \quad [5]$$

where $E_\pm^i = \frac{1}{2}(1 \pm \sigma_z^i)$ are idempotent operators. The Hadamard gate, $\hat{H} = \frac{1}{\sqrt{2}}(\sigma_x + \sigma_z)$, is a single qubit operation corresponding to a 180° rotation about the x=z axis in the X-Z plane.

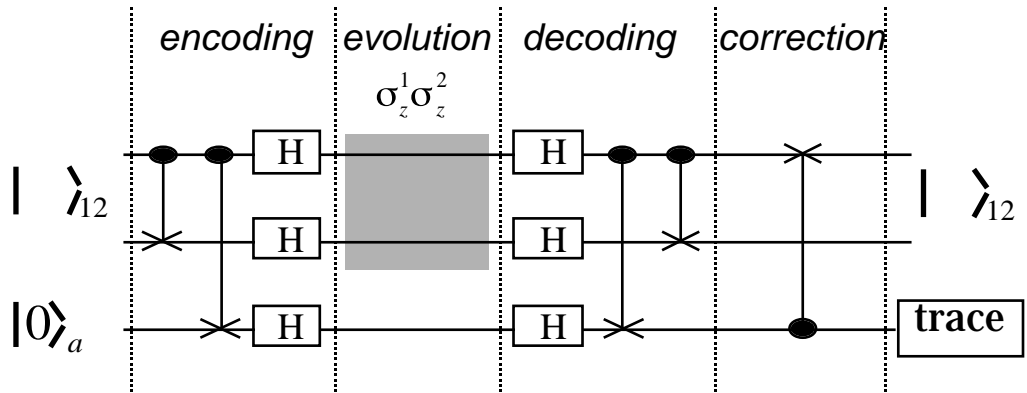

**Figure 1:** A code for protecting two data spin against evolution under $\sigma_z^1\sigma_z^2$ term in the Hamiltonian using a "perfect" ancilla.

The two data qubits in the code are initially in an arbitrary state $|\ \rangle_{12}$ while the ancillary qubit is prepared in the pure state $|0\rangle_a$. The initial state of the system can be written as

$$|\ \rangle_{12}|0\rangle_a = (\alpha|00\rangle + \beta|10\rangle + \gamma|01\rangle + \delta|11\rangle)|0\rangle_a. \quad [6]$$

Encoding proceeds by applying a sequence of two c-NOT gates, which entangles the state of the ancilla with the state of the data qubits, followed the application of Hadamard rotations on all three spins, yielding the following superposition

$$\begin{aligned}\frac{1}{2\sqrt{2}}[&(\alpha+\beta+\gamma+\delta)|000\rangle + \\ &(\alpha-\beta+\gamma-\delta)|100\rangle + \\ &(\alpha-\beta-\gamma+\delta)|010\rangle + \\ &(\alpha+\beta-\gamma-\delta)|110\rangle + \\ &(\alpha-\beta+\gamma-\delta)|001\rangle + \\ &(\alpha+\beta+\gamma+\delta)|101\rangle + \\ &(\alpha+\beta-\gamma-\delta)|011\rangle + \\ &(\alpha-\beta-\gamma+\delta)|111\rangle]\end{aligned} \quad [7]$$

The outcome of decoding (which is the inverse of encoding) is trivial in the case of no error because of the reversible nature of the unitary operations. In the presence of a $\sigma_z^1 \sigma_z^2$ error the state of the system after decoding is given by

$$(\alpha|10\rangle + \beta|00\rangle + \gamma|11\rangle + \delta|01\rangle)|1\rangle_a. \quad [8]$$

Thus the ancilla is no longer entangled with the data qubits. A comparison between this expression and the initial state (Eq. [6]) reveals that as desired the state of the system is mapped into an orthogonal subspace defined by the ancilla. Moreover, the ancilla being in the state $|1\rangle_a$ is the error syndrome. To recover the original state of qubits 1 and 2 all that need be done is to flip the first qubit conditional on the state of the ancilla being $|1\rangle_a$, so the final state is

$$\alpha|001\rangle + \beta|101\rangle + \gamma|011\rangle + \delta|111\rangle = |\ \rangle_{12}|1\rangle_a. \quad [9]$$

Analysis using the product operator formalism

It is useful to describe the dynamics of a weakly coupled spin system in terms of the product operator formalism, since it allows the evolution under the internal Hamiltonian and under RF excitations to be expressed as rotations in a three dimensional operator space (16, 17).

Although true pure states are not available in nuclear spin systems at room temperature, they can be prepared in pseudo pure states (18, 19, 20, 21, 22), which are defined as

$$\rho_{PP} = (1-\varepsilon)M_s + \varepsilon\rho_P, \qquad [10]$$

where $\varepsilon$ is a small constant, $\rho_P$ is a pure state, $M_s$ stands for the totally mixed state and the subscript $s = 2^N$ denotes the dimension of Hilbert space spanned by $M$.

The code in Fig. 1 requires the preparation of the ancilla in the (pseudo-) pure state $|0\rangle$. We identify the states $|0\rangle$ and $|1\rangle$ with the eigenstates of the density operator $E_+$ and $E_-$ ($E_\pm = \frac{1}{2}(1 \pm \sigma_z)$). A detailed description of the procedure by which the ancilla was prepared in a pseudo-pure is given in the Experimental section. Since any density operator can be expanded into a sum of pure states $|\ \rangle\langle\ |$, it follows from the linearity of quantum mechanics that the code will also work on any mixed state of the data qubits, i.e.

$$\rho_{init} = \sum_{i=1}^{16} a_i B_i^{data} \otimes E_+^a, \qquad [11]$$

where $\{B_i^{data}\}$ is a set of 16 product operators that forms a basis for the data spins (17). For example in the Cartesian basis they are given by

$$\{\sigma_{\eta_1}^1 \sigma_{\eta_2}^2\} \qquad [12]$$

where $\sigma_{\eta_k}^k \in \{1, \sigma_x^k, \sigma_y^k, \sigma_z^k\}$ and $\frac{1}{2}\sigma_{\eta_k}^k$ are the components of the angular momentum of the $k$-th spin ($k = 1, 2$).

The c-NOT in Eq. [5] can be written as a product of commuting exponential propagators (23), as follows:

$$\text{c-NOT}_{AB} = e^{-i\frac{\pi}{4}} e^{i\frac{\pi}{4}\sigma_x^B} e^{i\frac{\pi}{4}\sigma_z^A} e^{-i\frac{\pi}{4}\sigma_x^B \sigma_z^A}. \qquad [13]$$

Similarly, the Hadamard gate is given by

$$H = \frac{1}{\sqrt{2}}(\sigma_x + \sigma_z) = e^{i\frac{\pi}{2}\left(1 - \frac{1}{\sqrt{2}}(\sigma_x + \sigma_z)\right)} = ie^{-i\frac{\pi}{2}\sigma_x} e^{-i\frac{\pi}{4}\sigma_y}. \qquad [14]$$

Encoding is accomplished by successive application of c-NOT$_{1a}$ and c-NOT$_{2a}$ followed by Hadamard rotations on all three spins (see Fig. 1), so the encoding propagator is

$$e^{-i\frac{\pi}{2}\sigma_x^a} e^{-i\frac{\pi}{4}\sigma_y^a} e^{i\frac{\pi}{4}\sigma_x^a} e^{-i\frac{\pi}{2}\sigma_x^2} e^{-i\frac{\pi}{4}\sigma_y^2} e^{i\frac{\pi}{4}\sigma_x^2} e^{-i\frac{\pi}{2}\sigma_x^1} e^{-i\frac{\pi}{4}\sigma_y^1} e^{i\frac{\pi}{4}\sigma_z^1} e^{-i\frac{\pi}{4}\sigma_x^a \sigma_z^1} e^{-i\frac{\pi}{4}\sigma_x^a \sigma_z^2}. \qquad [15]$$

**Table:** Product operator description of state of the spin system evolving under a $\frac{1}{2}\varphi\sigma_z^1\sigma_z^2$ term in the Hamiltonian and protected by the code in Fig. 1.

| Initial† | Encoded | Decoded | Corrected‡ |
|---|---|---|---|
| $E_+^a$ | $\frac{1}{2}(1+\sigma_x^1\sigma_x^a)$ | $\frac{1}{2}(1+\sigma_z^a\cos(\varphi)-\sigma_x^1\sigma_y^a\sin(\varphi))$ | $E_\varphi^a$ |
| $\sigma_x^1 E_+^a$ | $\frac{1}{2}(\sigma_y^1\sigma_z^2\sigma_y^a - \sigma_z^1\sigma_z^2\sigma_z^a)$ | $\frac{1}{2}(\sigma_x^1 + \sigma_x^1\sigma_z^a\cos(\varphi) - \sigma_y^a\sin(\varphi))$ | $\sigma_x^1 E_\varphi^a$ |
| $\sigma_y^1 E_+^a$ | $\frac{1}{2}(\sigma_z^1\sigma_z^2\sigma_y^a + \sigma_y^1\sigma_z^2\sigma_z^a)$ | $\frac{1}{2}(\sigma_y^1\sigma_z^a + \sigma_y^1\cos(\varphi) + \sigma_z^1\sigma_x^a\sin(\varphi))$ | $\sigma_y^1 E_\varphi^a$ |
| $\sigma_z^1 E_+^a$ | $\frac{1}{2}(\sigma_x^1 + \sigma_x^a)$ | $\frac{1}{2}(\sigma_z^1\sigma_z^a + \sigma_z^1\cos(\varphi) - \sigma_y^1\sigma_x^a\sin(\varphi))$ | $\sigma_z^1 E_\varphi^a$ |
| $\sigma_x^2 E_+^a$ | $\frac{1}{2}(-\sigma_z^2 - \sigma_x^1\sigma_z^2\sigma_x^a)$ | $\frac{1}{2}(\sigma_x^2 + \sigma_x^2\sigma_z^a\cos(\varphi) - \sigma_x^1\sigma_x^2\sigma_y^a\sin(\varphi))$ | $\sigma_x^2 E_\varphi^a$ |
| $\sigma_x^1\sigma_x^2 E_+^a$ | $\frac{1}{2}(-\sigma_y^1\sigma_y^a + \sigma_z^1\sigma_z^a)$ | $\frac{1}{2}(\sigma_x^1\sigma_x^2 + \sigma_x^1\sigma_x^2\sigma_z^a\cos(\varphi) - \sigma_x^2\sigma_y^a\sin(\varphi))$ | $\sigma_x^1\sigma_x^2 E_\varphi^a$ |
| $\sigma_y^1\sigma_x^2 E_+^a$ | $\frac{1}{2}(-\sigma_z^1\sigma_y^a - \sigma_y^1\sigma_z^a)$ | $\frac{1}{2}(\sigma_y^1\sigma_x^2\sigma_z^a + \sigma_y^1\sigma_x^2\cos(\varphi) + \sigma_z^1\sigma_x^2\sigma_x^a\sin(\varphi))$ | $\sigma_y^1\sigma_x^2 E_\varphi^a$ |
| $\sigma_z^1\sigma_x^2 E_+^a$ | $\frac{1}{2}(-\sigma_x^1\sigma_z^2 - \sigma_z^2\sigma_x^a)$ | $\frac{1}{2}(\sigma_z^1\sigma_x^2\sigma_z^a + \sigma_z^1\sigma_x^2\cos(\varphi) - \sigma_y^1\sigma_x^2\sigma_x^a\sin(\varphi))$ | $\sigma_z^1\sigma_x^2 E_\varphi^a$ |
| $\sigma_y^2 E_+^a$ | $\frac{1}{2}(\sigma_x^1\sigma_y^2 + \sigma_y^2\sigma_x^a)$ | $\frac{1}{2}(\sigma_y^2 + \sigma_y^2\sigma_z^a\cos(\varphi) - \sigma_x^1\sigma_y^2\sigma_y^a\sin(\varphi))$ | $\sigma_y^2 E_\varphi^a$ |
| $\sigma_x^1\sigma_y^2 E_+^a$ | $\frac{1}{2}(-\sigma_z^1\sigma_x^2\sigma_y^a - \sigma_y^1\sigma_x^2\sigma_z^a)$ | $\frac{1}{2}(\sigma_x^1\sigma_y^2 + \sigma_x^1\sigma_y^2\sigma_z^a\cos(\varphi) - \sigma_y^2\sigma_y^a\sin(\varphi))$ | $\sigma_x^1\sigma_y^2 E_\varphi^a$ |
| $\sigma_y^1\sigma_y^2 E_+^a$ | $\frac{1}{2}(\sigma_y^1\sigma_x^2\sigma_y^a - \sigma_z^1\sigma_x^2\sigma_z^a)$ | $\frac{1}{2}(\sigma_y^1\sigma_y^2\sigma_z^a + \sigma_y^1\sigma_y^2\cos(\varphi) + \sigma_z^1\sigma_y^2\sigma_x^a\sin(\varphi))$ | $\sigma_y^1\sigma_y^2 E_\varphi^a$ |
| $\sigma_z^1\sigma_y^2 E_+^a$ | $\frac{1}{2}(\sigma_y^2 + \sigma_x^1\sigma_y^2\sigma_x^a)$ | $\frac{1}{2}(\sigma_z^1\sigma_y^2\sigma_z^a + \sigma_z^1\sigma_y^2\cos(\varphi) - \sigma_y^1\sigma_y^2\sigma_x^a\sin(\varphi))$ | $\sigma_z^1\sigma_y^2 E_\varphi^a$ |
| $\sigma_z^2 E_+^a$ | $\frac{1}{2}(\sigma_x^1\sigma_x^2 + \sigma_x^2\sigma_x^a)$ | $\frac{1}{2}(\sigma_z^2 + \sigma_z^2\sigma_z^a\cos(\varphi) - \sigma_x^1\sigma_z^2\sigma_y^a\sin(\varphi))$ | $\sigma_z^2 E_\varphi^a$ |
| $\sigma_x^1\sigma_z^2 E_+^a$ | $\frac{1}{2}(\sigma_z^1\sigma_y^2\sigma_y^a + \sigma_y^1\sigma_y^2\sigma_z^a)$ | $\frac{1}{2}(\sigma_x^1\sigma_z^2 + \sigma_x^1\sigma_z^2\sigma_z^a\cos(\varphi) - \sigma_z^2\sigma_y^a\sin(\varphi))$ | $\sigma_x^1\sigma_z^2 E_\varphi^a$ |
| $\sigma_y^1\sigma_z^2 E_+^a$ | $\frac{1}{2}(-\sigma_y^1\sigma_y^2\sigma_y^a + \sigma_z^1\sigma_y^2\sigma_z^a)$ | $\frac{1}{2}(\sigma_y^1\sigma_z^2\sigma_z^a + \sigma_y^1\sigma_z^2\cos(\varphi) + \sigma_z^1\sigma_z^2\sigma_x^a\sin(\varphi))$ | $\sigma_y^1\sigma_z^2 E_\varphi^a$ |
| $\sigma_z^1\sigma_z^2 E_+^a$ | $\frac{1}{2}(\sigma_x^2 + \sigma_x^1\sigma_x^2\sigma_x^a)$ | $\frac{1}{2}(\sigma_z^1\sigma_z^2\sigma_z^a + \sigma_z^1\sigma_z^2\cos(\varphi) - \sigma_y^1\sigma_z^2\sigma_x^a\sin(\varphi))$ | $\sigma_z^1\sigma_z^2 E_\varphi^a$ |

† $E_+^a \quad \frac{1}{2}(1+\sigma_z^a)$

‡ $E_\varphi^a \quad \frac{1}{2}(1+\sigma_z^a\cos(\varphi)-\sigma_y^a\sin(\varphi)) = e^{-i\frac{\varphi}{2}\sigma_x^a} E_+^a e^{i\frac{\varphi}{2}\sigma_x^a}$

Note that $Tr_a[B_i^{data} E_+^a] = Tr_a[B_i^{data} E_\varphi^a] = B_i^{data}$

The resultant encoded state for the set of input product operators $B_i^{data} E_+^a$ (in Eq. [11]) is given in the Table. In case of no error the original state is trivially recovered by the decoding sequence. However in the more interesting case of evolution under a $\frac{1}{2}\sigma_z^1 \sigma_z^2$ coupling term by an angle $\varphi$, the state of the two data spins is recovered after the correction step. The density operators for the decoded state as well as the corrected state are listed in the Table.

The table shows that following a trace over the ancilla all basis states of the data spins are fully recovered, indicating that the code will correct any initial state, as desired. Furthermore, a comparison between the initial and the final states reveals that phase evolution of the data spins due to evolution under coupling is mapped by the code to a rotation about the *x*-axis of the ancilla. This effect, expressed symbolically by $E_z^a \xrightarrow{\frac{1}{2}\varphi \sigma_z^1 \sigma_z^2} E_\varphi^a \equiv e^{-i\frac{\varphi}{2}\sigma_x^a} E_+^a e^{i\frac{\varphi}{2}\sigma_x^a}$, implies that the transverse component of the ancilla is informative of the phase evolution $\varphi$ due to the error $\sigma_z \sigma_z$. However, any information held by the ancilla is erased by the trace at the end (see Fig. 1). The trace over the ancilla is obtained be decoupling it during acquisition of the spectra.

## EXPERIMENTAL AND METHODS

The three bit quantum code in Fig. 1 was realized in a sample of $^{13}$C labeled alanine $\left(NH_3^+ - C^\alpha H(C^\beta H_3) - C\, O_2^-\right)$ in D$_2$O at room temperature. Measurements were conducted on a Bruker AMX400 spectrometer (9.6 T) equipped with a 5 mm probe tuned to $^{13}$C and $^1$H frequencies of 100.61 MHz and 400.13 MHz, respectively. The probe was equipped with xyz-gradient coils capable of generating field gradients ranging from -60 to +60 G/cm.

With decoupling of the protons alanine exhibits a weakly-coupled carbon spectrum. The internal Hamiltonian of this system in the rotating frame is given by Eq. [3] where the chemical shifts, coupling constants (assuming the transmitter is set on $C^\alpha$) and relaxation times are:

$$\omega^C/(2\pi) = 12580\,Hz \qquad \omega^{C^\alpha}/(2\pi) = 0\,Hz \qquad \omega^{C^\beta}/(2\pi) = -3443\,Hz$$
$$J^{CC^\alpha} = 54.2\,Hz \qquad J^{CC^\beta} = 35.1\,Hz \qquad J^{C^\alpha C^\beta} = 1.2\,Hz$$
$$T_1^C = 20.1\,s \qquad T_1^{C^\alpha} = 2.5\,s \qquad T_1^{C^\beta} = 1.6\,s$$
$$T_2^C = 0.45\,s \qquad T_2^{C^\alpha} = 0.42\,s \qquad T_2^{C^\beta} = 0.8\,s. \qquad [16]$$

The $C^\alpha$ and the carbonyl $C$ were chosen as data spin #1 and data spin #2, respectively, while $C^\beta$ was used as the ancilla.

The logic gates (c-NOT, Hadamard) were implemented using time delays and selective RF excitations. The latter included phase-modulated Gaussian shaped pulses (24), in addition to rectangular and composite soft pulses. A detailed description of the basic modules used for implementing the logic network can be found in (8).

Preparation of the initial states

Starting from thermal equilibrium, the (deviation part of the) density operator $\rho_{eq} = \tfrac{1}{2}\sigma_z^1 + \tfrac{1}{2}\sigma_z^2 + \tfrac{1}{2}\sigma_z^a$ describes the initial state of the spin system. The data spins were prepared in an arbitrary state with the third spin (ancilla) in a pseudopure state $E_+^a = \tfrac{1}{2}(1+\sigma_z^a)$. To prepare the state $\sigma_x^1 E_+^a$ a selective $\pi/2$ pulse was applied to spin 2 followed by a crusher magnetic field gradient, yielding the density operator $\tfrac{1}{2}\sigma_z^1 + \tfrac{1}{2}\sigma_z^a$. Application of the pulse sequence

$$\frac{\pi}{2}\bigg|_{-x}^{a} - \frac{1}{2J_{1a}} - \frac{\pi}{2}\bigg|_{y}^{a} - \frac{\pi}{2}\bigg|_{y}^{1} \qquad [17]$$

then led to the desired state $\sigma_x^1(1+\sigma_z^a) = \sigma_x^1 E_+^a$, where $1/(2J_{1a})$ is an evolution under the effective Hamiltonian $\tfrac{\pi}{2} J_{1a}\sigma_z^1\sigma_z^a$ for a time $1/(2J_{1a})$.

Similarly, to create the state $\sigma_x^1\sigma_z^2 E_+^a$ the following sequence was applied to the state $\tfrac{1}{2}\sigma_z^1 + \tfrac{1}{2}\sigma_z^a$

$$\frac{\pi}{2}\bigg|_{-x}^{a} - \frac{1}{2J_{1a}} - \frac{\pi}{2}\bigg|_{y}^{a} - \frac{\pi}{2}\bigg|_{x}^{1} - \frac{1}{2J_{12}}. \qquad [18]$$

Encoding, decoding and correction

The encoding propagator in Eq. [15] was implemented by the sequence

$$\frac{\pi}{2}\bigg|_{x}^{\{2,a\}} - \frac{\pi}{2}\bigg|_{y}^{\{2,a\}} - \frac{1}{2J_{12}} - \frac{1}{2J_{1a}} - \frac{\pi}{2}\bigg|_{y}^{1} - (\pi)_x^{\{2,a\}} \qquad [19]$$

where $(\theta)_\varphi^{\{k,l\}}$ stands for a selective $\theta$ pulse about the $\varphi$ axis applied to both the $k$ and $l$ spins simultaneously. Accordingly, the decoding sequence was implemented by

$$(\pi)_x^{\{2,a\}} - \frac{\pi}{2}\bigg|_{-y}^{1} - \frac{1}{2J_{12}} - \frac{1}{2J_{1a}} - \frac{\pi}{2}\bigg|_{-y}^{\{2,a\}} - \frac{\pi}{2}\bigg|_{-x}^{\{2,a\}}. \qquad [20]$$

The correction step is followed immediately by a partial trace over the ancilla, i.e. by observing the data spins while decoupling the ancilla. Therefore correction propagators that operate only on the ancilla can be eliminated with no effect on the final result. Since the net phase in Eq. [13] is also unobservable, this leaves only the propagator $e^{i\frac{\pi}{4}\sigma_x^1}e^{-i\frac{\pi}{4}\sigma_x^1\sigma_z^a}$, which was implemented by the sequence

$$\frac{\pi}{2}\bigg|_{-y}^{1} - \frac{1}{2J_{1a}} - \frac{\pi}{2}\bigg|_{y}^{1} - \frac{\pi}{2}\bigg|_{-x}^{1}. \qquad [21]$$

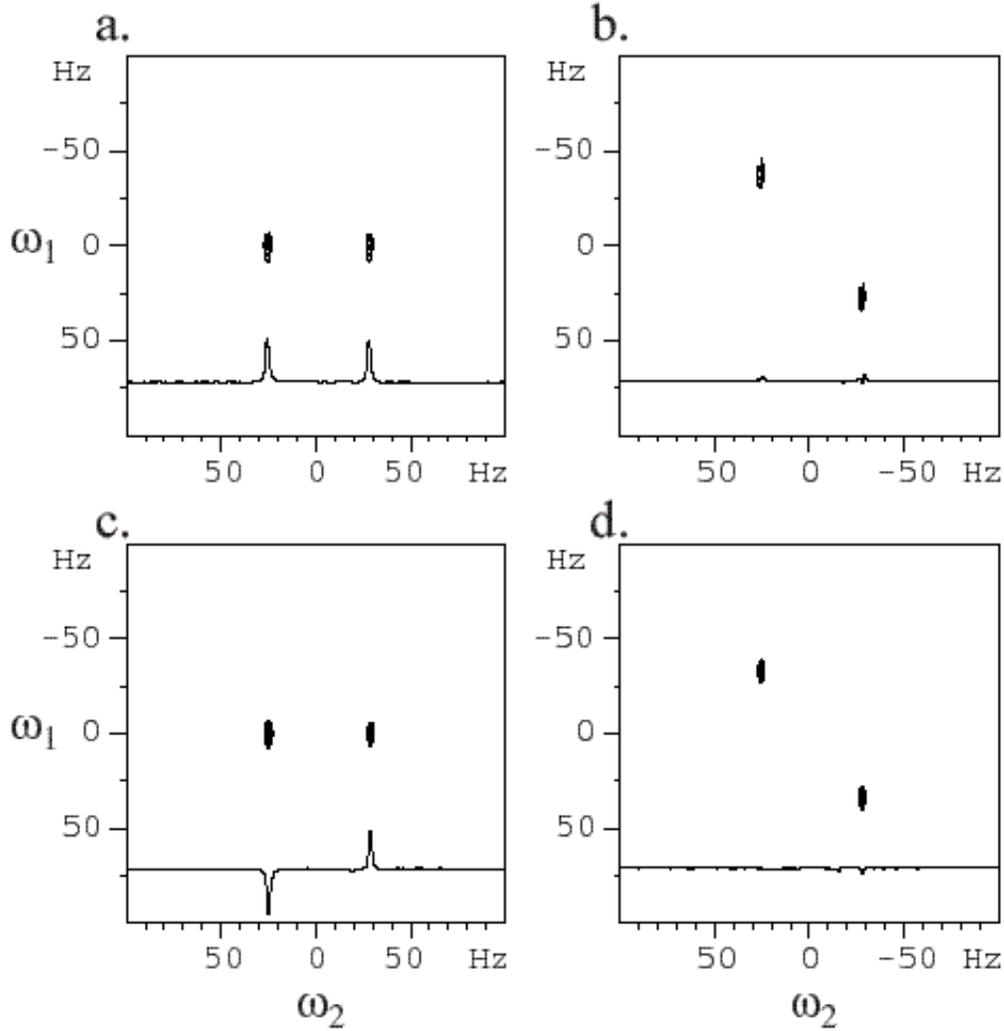

**Figure 2:** Protecting two data spins against evolution under *J*-coupling using the code in Fig.1. 2D carbon spectra of the 1$^{st}$ data spin (C ) are shown for two initial states of the data spins with correction (left panels) and without correction (right panels). In (a) and (b) the data spins are initially prepared in the state $\sigma_x^1$, while in (c) and (d) their initial state is $\sigma_x^1 \sigma_z^2$ (see text). In both cases the third spin, the ancilla, is initially prepared in a pseudo-pure state $E_+^a$. Cross-peaks due to evolution under scalar coupling, apparent only in the uncorrected spectra ($\omega_1 = \pm 27 Hz$), are completely suppressed by the code. 1D spectra corresponding to $\omega_1 = 0$ are plotted at the bottom of each panel. The two peaks in the corrected spectra are (a) in-phase and (c) in anti-phase, in accordance with the state of the data spin.

## RESULTS AND DISCUSSION

The quantum code shown in Fig. 1 was applied to the states $\sigma_x^1 E_+^a$ and $\sigma_x^1 \sigma_z^2 E_+^a$. The spectrum of the first data spin ($C^\alpha$) was recorded as a function of the evolution time $\tau$ under the effective Hamiltonian $\frac{\pi}{2} J_{12} \sigma_z^1 \sigma_z^2$, which was obtained via the pulse sequence

$$-\frac{\tau}{8} - (\pi)_y^a - \frac{\tau}{8} - (\pi)_y^{\{1,2,a\}} - \frac{\tau}{8} - (\pi)_{-y}^a - \frac{\tau}{4} - (\pi)_y^a - \frac{\tau}{8} - (\pi)_{-y}^{\{1,2,a\}} - \frac{\tau}{8} - (\pi)_{-y}^a - \frac{\tau}{8} - \quad [22]$$

where $(\pi)^{\{1,2,a\}}$ denotes a non-selective $\pi$ rotation. Note that no selective pulses were employed on either of the data spins. In the experiments $\tau$ was incremented in 32 steps of $1/(16 J_{12}) = 1/(16 \times 54.2) = 1.1531 m\sec$. Two-dimensional (2D) spectra were obtained via Fourier transformations in respect to $\tau$ ($\omega_1$) and the acquisition time ($\omega_2$).

In Fig. 2, these two-dimensional spectra of the first data spin are plotted with and without quantum coding for the two initial states. The apparent cross peaks (at $\omega_1/2\pi = \pm 27 Hz$) in the uncorrected spectra indicate phase evolution due to the scalar coupling, $\frac{\pi}{2} J_{12} \sigma_z^1 \sigma_z^2$, between the two data spins. Conversely, their disappearance in the error corrected spectra indicates a complete suppression of this coupling by the code. Error corrected spectra obtained for the initial state $\sigma_x^1 E_+^a$ exhibit two in-phase peaks, whereas the $\sigma_x^1 \sigma_z^2 E_+^a$ exhibit two anti-phase peaks, as expected from the state of the data spins.

The code implemented in this study (see Fig.1) requires the preparation of the ancilla in a (pseudo-) pure state. However, this is not a general requirement. In fact, a symmetric version of this code is capable of protecting the state of the two data qubits with one isolated ancilla in an arbitrary state (see Fig. 3).

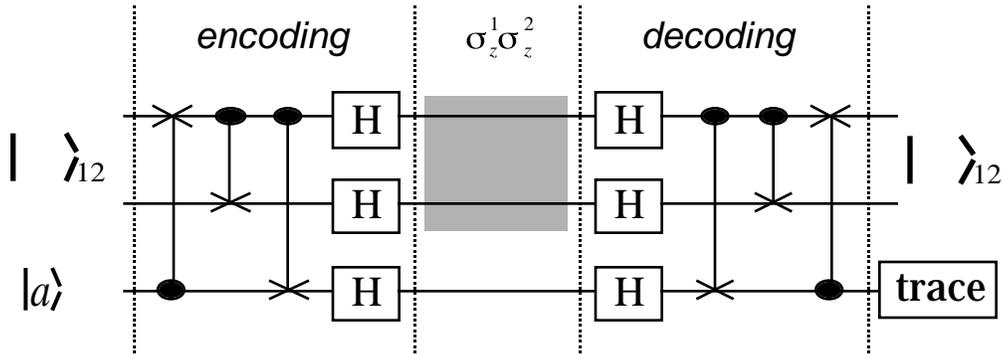

**Figure 3:** A symmetric version of the code in Fig. 1 that protects two data spins against evolution under a $\sigma_z^1 \sigma_z^2$ term in the Hamiltonian using a "perfect" ancilla. Here, however, the ancilla can be in arbitrary state so that no preparation of a pseudo-pure state is required.

In the presence of a $\sigma_z^1 \sigma_z^2$ error, i.e. evolution under $J_{12}$-coupling, the initial state of the system given by

$$\begin{aligned} \rho_i &= |\ \rangle_{12} |a\rangle \\ &= |\ \rangle_{12} (\alpha_a |0\rangle_a + \beta_a |1\rangle_a) \end{aligned} \qquad [23]$$

is mapped by the error correction code to the state

$$\begin{aligned} \rho_f &= |\ \rangle_{12} (\beta_a |0\rangle_a + \alpha_a |1\rangle_a) \\ &= |\ \rangle_{12} (\sigma_x^a |a\rangle) \end{aligned} . \qquad [24]$$

In other words, the $\sigma_z^1 \sigma_z^2$ error is mapped to a bit flip error ($\sigma_x^a$) on the ancilla. The dispensable feature of the ancilla as well as its isolation from phase distortions during the error period plays an important role in the generation of this code. It allows a true mapping of the error propagator from the data channels to the dispensable channel, i.e. the ancilla. That this is unconditional on the state of the latter is of great importance since it saves elaborate preparation procedures and further loss in polarization (6, 25). Interestingly, it suggests that, at least for this case, encoding in an isolated and dispensable channel may be found by means of average Hamiltonian theory. Thus, the generator for the error $\tfrac{1}{2}\varphi \sigma_z^1 \sigma_z^2$ can be mapped to a single rotation on the ancilla, for example $\tfrac{1}{2}\varphi \sigma_x^a$, by

$$e^{i\tfrac{1}{2}\varphi \sigma_x^a} = \ e^{i\tfrac{1}{2}\varphi \sigma_z^1 \sigma_z^2}\ {}^{-1} \qquad [25]$$

where the unitary transformation can be constructed from available rotation generators in the Hamiltonian (26) . In the present case, the mapping propagator

$$= e^{i\tfrac{\pi}{4}\sigma_z^1 \sigma_z^a} e^{i\tfrac{\pi}{4}\sigma_y^a} e^{i\tfrac{\pi}{4}\sigma_y^1} e^{i\tfrac{\pi}{4}\sigma_z^1 \sigma_z^a} e^{i\tfrac{\pi}{4}\sigma_z^1 \sigma_z^2} e^{-i\tfrac{\pi}{4}\sigma_y^1} \qquad [26]$$

is equivalent to the encoding in Fig. 3. Similarly, any set of $N_e$ products of Pauli operators including the identity, which all commute with one another and act only on the data qubits, can be mapped unitarily to another set of operators acting exclusively on $N_a$ isolated ancilla qubits, provided that $N_a \ \log_2 N_e$.

The ability to generate a code for protecting against certain errors without the requirement for special ancilla preparation is a privilege saved for the case of isolated qubits. In more general and realistic cases where errors can, and do, occur on each one of the qubits including the ancilla, the preparation of the ancilla is essential. Consider for example the repetition code for correcting against single qubit phase errors, which was demonstrated in our previous studies (6, 8) and is given in Fig. 4. The code corrects for phase errors occurring on any one of the three qubits. Therefore it corrects for the coherent evolution of one spin under a $\sigma_z$ term in the Hamiltonian, or alternatively it corrects to first order for coherent evolution under a sum of $\sigma_z$'s for all three spins. In this example all three qubits are subjected to errors, no qubit is isolated, and therefore the information on the data qubit must be protected in a superposition of the three qubits' state. Hence the preparation of ancillary spins in a pseudo-pure state is essential because mapping of the set of single phase errors $\{\sigma_z^{obs},\sigma_z^{a1},\sigma_z^{a2}\}$ to a set of errors consisting only of operators acting on the ancilla is not feasible.

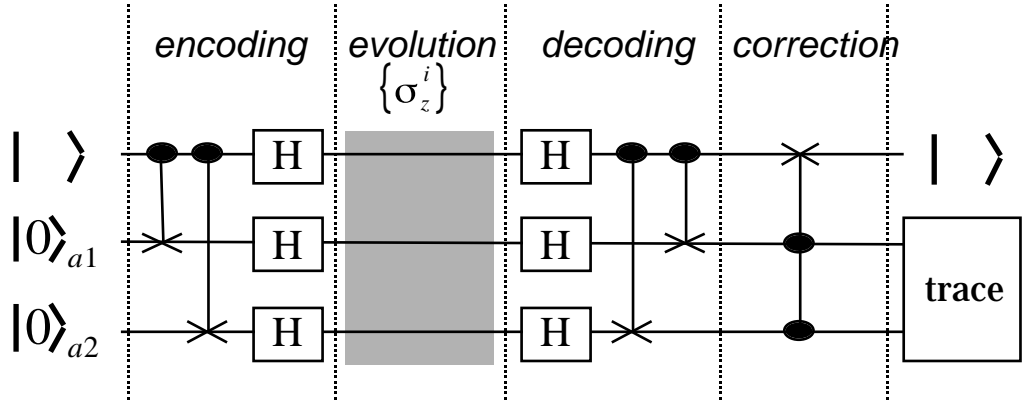

**Figure 4:** A quantum error correction code for protecting one date spin against single qubit phase errors.

## QUANTUM CODES AND GENERAL EVOLUTION UNDER WEAK COUPLING $\sigma_z\sigma_z$ TERMS

In the context of quantum information processing bilinear product errors can be handled by the concatenation of codes designed to address single particle errors. The only requirement regarding the type of physical error, coherent or incoherent, is that the code must be run sufficiently quickly so as to reliably correct for the error. One very useful observation is that physically these errors can only arise through 1 qubit and 2 qubit processes. Therefore it need not concern us that a $n$-qubit coherence decays between $n$ and $n^2$ times more rapidly than a single qubit error, as long as the code (or concatenated code) is able to handle both one and two qubit errors. Hence the bilinear product terms must be considered along with the single products, but no higher order terms need be dealt with.

As we have shown, evolution of a spin system under a $\sigma_z\sigma_z$ term in the Hamiltonian can be recovered by a quantum code with the addition of a single isolated ancilla. In this part we would like to consider the case of non-isolated ancillae in which all bilinear couplings terms in the Hamiltonian are active. Thus, the Hamiltonian of the system during that period is given by

$$H_{\text{int}} = \sum_{k<l}^{N} \tfrac{\pi}{2} J_{kl} \sigma_z^k \sigma_z^l \qquad [27]$$

where $N$ is the total number of spins in the system ($N$ = number of data qubits + number of ancilla), and $J_{kl}$ are coupling constants which in general vary considerably. In NMR, interactions bilinear in the spin operators originate from indirect, electron mediated scalar couplings as well as from direct dipolar interactions. Scalar couplings decrease with the number of bonds between the atoms. Dipolar couplings between the nuclear magnetic moments fall off with distance as $1/r^3$ and in the presence of high external magnetic field their angular dependency follows $\tfrac{1}{2}(3\cos^2\theta - 1)$, where $\theta$ is the angle between the internuclear vector and the magnetic field.

The spin system evolves under the Hamiltonian in Eq. [27] with the propagator

$$\begin{aligned} U(t) &= \prod_{k<l}^{N} e^{-i\tfrac{\pi}{2} J_{kl} \sigma_z^k \sigma_z^l t} \\ &= \prod_{k<l}^{N} \left(\cos(\pi J_{kl} t/2) - i\sigma_z^k \sigma_z^l \sin(\pi J_{kl} t/2)\right) \end{aligned} \qquad [28]$$

In the following we provide simple arguments to show that adding ancilla spins does not provide a large enough encoding space (number of orthogonal subspaces) to accommodate all possible errors. Each addition of an ancilla spin (I=1/2) doubles the encoding capacity, which therefore grows exponentially as $2^{N_a}$ with the number of ancillae $N_a$. However, adding the $n$-th ancilla spin also introduces an additional $n-1$ coupling terms into the sum in Eq. [27]. Accordingly, the total number of different product operator terms obtained by expanding Eq. [28], taking into account the relation $\left(\sigma_z^k\right)^2 = 1$ and including the identity operator for no error, is given by

$$\sum_{m=0}^{N/2} \binom{N}{2m} = 2^{N-1} \qquad [29]$$

where $m$ is the order of the error, i.e. $m=0$ corresponds to no error; $m=1$ to $\sigma_z^k \sigma_z^l$ ($k \neq l$) errors; $m=2$ to $\sigma_z^k \sigma_z^l \sigma_z^u \sigma_z^v$ ($k \neq l \neq u \neq v$) errors, etc.

A single data spin can be protected with any number of ancilla since $N_a = N - 1$, however errors will completely occupy the encoding space regardless of the number. An example for such a code with a single ancilla is given in Fig. 5. The system initially prepared in the state

$$(\alpha|0\rangle + \beta|1\rangle)|0\rangle_a = \alpha|00\rangle + \beta|10\rangle \qquad [30]$$

is encoded by the Hadamard transformation to the superposition

$$\tfrac{1}{2}(\alpha+\beta)|00\rangle + \tfrac{1}{2}(\alpha-\beta)|10\rangle + \tfrac{1}{2}(\alpha+\beta)|10\rangle + \tfrac{1}{2}(\alpha-\beta)|11\rangle \qquad [31]$$

In the case of no error the original state Eq. [30] is trivially recovered by the decoding Hadamard transformation. In the presence of a $\sigma_z^1 \sigma_z^2$ error, on the other hand, the state after decoding,

$$\alpha|11\rangle + \beta|01\rangle = (\alpha|1\rangle + \beta|0\rangle)|1\rangle_a, \qquad [32]$$

is corrected by flipping the first qubit conditional on the second being in the state $|1\rangle$.

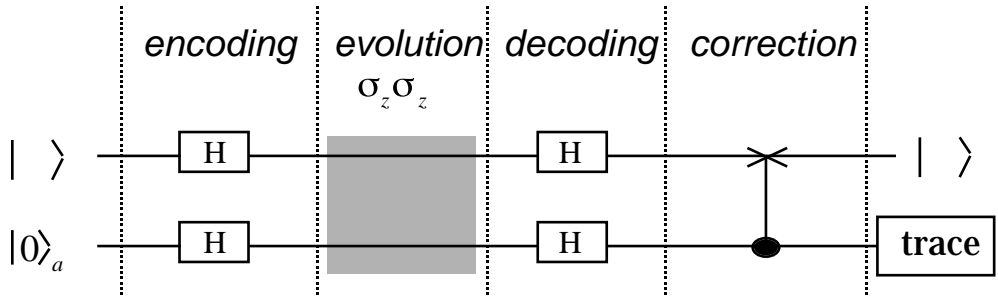

**Figure 5:** A simple code for protecting a single date spin against evolution under $\sigma_z \sigma_z$ term in the Hamiltonian of a two spin system, using a single ancilla.

For any system with more than one data spin the total number of errors will exceed the encoding capacity by a factor of $2^{N-N_a-1}$. Nevertheless, the execution of any quantum logic operation is confined in time and therefore for most purposes the requirement for complete control can be traded with a fault tolerant approach. In other words, terms in the expansion of Eq. [28] that require much longer time (compared to some threshold) to evolve than the time required to execute an elementary gate need not be corrected (27).

Fault-tolerant error correction is designed to deal with incoherent, independent errors on single qubits. As multiple single qubit errors begin to accumulate, higher-order multilinear operators will become significant (just as operators of order greater than two become significant in Eq. [28] at longer times). As mentioned earlier, such multilinear operators can be dealt with by concatenating codes for single qubit errors, which can in principle enable arbitrarily precise control of decoherence, providing the error rate is below some threshold (28). Concatenation is also able to handle bilinear error operators which arise directly through coherent interactions between pairs of qubits, but requires many more ancilla and gates than do codes which are designed specifically for bilinear errors. For instance if one is interested in protecting against a finite set of direct bilinear product errors without restoring single particle errors then other, less demanding, schemes can be used.

Returning to the evolution of a system under the Hamiltonian of Eq. [27], it may be seen from Eq. [28] that for a sufficiently short time (such that $\max|J_{kl}|t <<\pi/2$) the effect of the higher-order terms can be neglected. The maximum correctable order $m_{max}$ can be determined from the condition

$$\sum_{m=0}^{m_{max}} \binom{N}{2m} \leq 2^{N_a} \qquad [33]$$

which in the case of two data spins ($N = N_a + 2$) simplifies to

$$N_a \geq 2m_{max}. \qquad [34]$$

Thus the minimum number of ancilla qubits required to protect two data spins to first order against any one $\sigma_z\sigma_z$ evolution is four. An example of such a code is given in Fig.6. Unlike the code in Fig.1 this code corrects for a total of $\binom{6}{2}=15$ possible errors on all $\sigma_z^i\sigma_z^j$ coupling pairs, therefore the number of subspaces required for encoding is 16 (an additional subspace is required for no error).

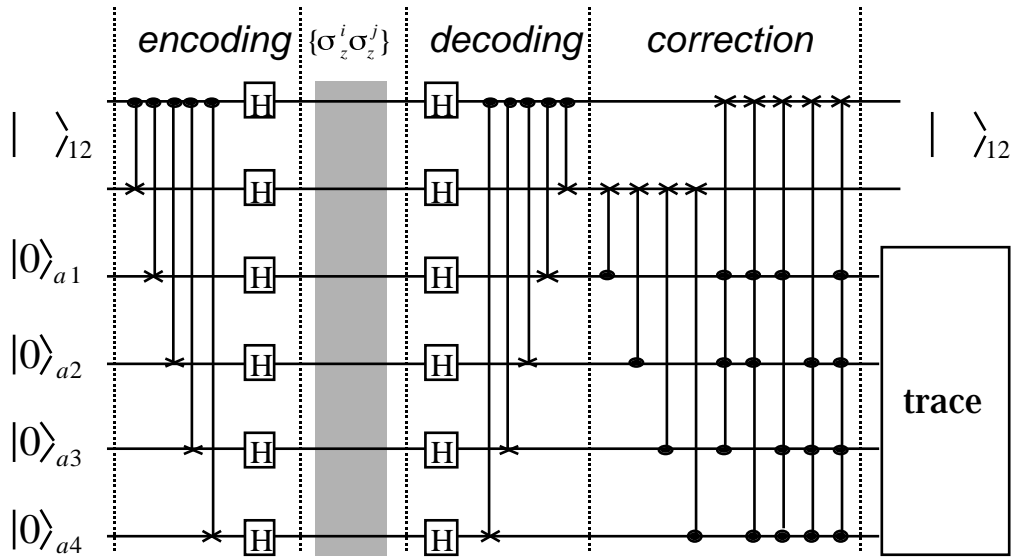

**Figure 6:** A quantum code for protecting two data spins to first order against evolution under $\sigma_z\sigma_z$ terms using four ancillae.

## CONCLUSIONS

Quantum coding is a powerful method capable of protecting quantum states from both coherent as well as incoherent evolution, at the expense of additional ancilla spins. It is complementary to existing methods of coherent averaging which can unitarily transform and reduce the magnitude of terms in the Hamiltonian of the system of interest, in that it can replace and transform these terms by other, perhaps more desirable terms involving the ancillae. This may be particularly useful when the frequencies in the system of interest are unknown, or degeneracies prevent it from being directly manipulated as desired. Although error correcting codes were devised to control incoherent errors, quantum coding also provides a means of exploring both coherent and incoherent processes, because the failure of a given code to control a system indicates that other dynamical processes are involved. It is anticipated that ideas from quantum coding will play an increasingly important role in the design of NMR experiments for solving a variety of practical problems.

## ACKNOWLEDGEMENTS

We thank Ching-Hua Tseng for helpful discussions. This work was supported by the U.S. Army Research Office under grant number DAAG-55-97-1-0342 from the DARPA Microsystems Technology Office.

## REFERENCES

1. P. Shor, Phys. Rev. A **52**, 2493 (1995).
2. A. M. Steane, Phys. Rev. Lett. 77, 793-797 (1996).
3. R. Laflamme, C. Miquel, J. P. Paz, W. Zurek, Phys. Rev. Lett. 77, 198 (1996).
4. J. Preskill, Proc. Roy. Soc. London, Ser. A 454, 385 (1998).
5. E. Knill and R. Laflamme, Phys. Rev. A 55, 900 (1997).
6. D. G. Cory, M.D. Price, W. Maas, E. Knill, R. Laflamme, W. H. Zurek, T. F. Havel, and S. S. Somaroo, Phys. Rev. Lett. 81, 2152 (1998).
7. D. W. Leung, L. M. K. Vandersypen, X. Zhou, M. H. Sherwood, C. S. Yannoni, M. G. Kubinec, I. L. Chuang, quant-ph/9811068.
8. Y. Sharf, D. G. Cory, S. S. Somaroo, E. Knill, R. Laflamme, W. H. Zurek, and T. F. Havel, Mol. Phys. (submitted).
9. M. B. Ruskai, quant-ph/9906114.
10. D. A. Lidar, D. Bacon, J. Kempe, and K. B. Whaley, quant-ph/9907096.
11. L. Viola, E. Knill, and S. Lloyd, quant-ph/0002072.
12. R. R. Ernst, G. Bodenhausen, and A. Wokaun, Principles of Nuclear Magnetic Resonance in One and Two Dimensions (Clarendon, Oxford, 1987).
13. A. Barenco , C. H. Bennett, R. Cleve, D. P. DiVincenzo, N.Margolus, P. Shor, T. Sleator, J. Smolin, H. Weinfurter, Phys. Rev. A 52, (5) 3457 (1995).
14. U. Haeberlen and J. Waugh, Phys. Rev. 175, 453 (1968).

15. D. G. Cory, J. B. Miller, and A. N. Garroway, J. Magn. Reson. 90, 205 (1990).
16. S. S. Somaroo, D. G. Cory, and T. F. Havel, Phys. Lett. A. 240, 1 (1998).
17. O. W. Sörensen, G. W. Eich, M. H. Levitt, G. Bodenhausen, and R. R. Ernst, Prog. NMR Spect. 16, 163 (1983).
18. D. G. Cory, A. F. Fahmy, and T. F. Havel, Pro. Natl. Acad. Sci. 94, 1634 (1997).
19. N. A. Gershenfeld and I. L. Chuang, Science 275, 350 (1997).
20. D. G. Cory, M. D. Price, and T. F. Havel, Physica D 120, 82 (1998).
21. E. Knill I. Chuang, and R. Laflamme, Phys. Rev. A 57, 3348 (1998).
22. I. L. Chuang, N. Gershenfeld, M. G. Kubinec, and D. W. Leung, Proc. R. Soc. Lond. A 454, 447 (1998).
23. M. D. Price, S. S. Somaroo, C-H Tseng, J. C. Gore, A. F. Fahmy, T. F. Havel, and D. G. Cory, J. Magn. Reson. 140, 371 (1999).
24. S. L. Patt, J. Magn. Reson. 96, 94 (1992).
25. W. S. Warren, Science, 277, 1688 (1997).
26. C. H. Tseng, S. Somaroo, T. F. Havel, K. Knill, R. Laflamme, Y. Sharf, and D. G. Cory, (quant-ph/9908012)
27. D. Gottesman, Phys. Rev. A 57, 127 (1998).
28. E. Knill, R. Laflamme, and W. H. Zurek, Proc. R. Soc. Lond. A 454, 365 (1998).